# Diameter Optimization for Highest Degree of Ballisticity of Carbon Nanotube Field Effect Transistors

*I. Khan, O. Morshed and S. M. Mominuzzaman*

*Department of Electrical and Electronic Engineering, Bangladesh University of Engineering and Technology, Dhaka-1000, Bangladesh*
*E-mail: imtiajkhan203@gmail.com*

Carbon Nanotube (CNT) is one of the most significant materials for the development of faster and improved performance of nano-scaled transistors. This paper aims at analyzing a trade-off between device performance and device size of CNT based transistor. Acoustic and Optical Phonon scattering along with the elastic scattering lead to the non-ballistic performances of those transistors. The main focus of this work is mainly on finding an optimum diameter to obtain the highest degree of ballisticity from both single-walled and double-walled Carbon Nanotube Field Effect Transistors (CNT-FETs). At first, an n-type single-walled Carbon Nanotube Field Effect Transistor has been considered and the diameter dependence on degree of ballisticity has been simulated. The effects of drain voltage, gate voltage and channel length have been investigated for such characteristics followed by a comparison with double-walled CNT-FET structure. First of all, it has been found that degree of ballisticity along with the optimum diameter increases with the increase of ($V_{DS} - V_{GS}$). The increase of channel length, however, degrades the ballistic performance demanding a higher diameter to reach the optimum point. Finally, it can be concluded that optimum diameter for DWCNT-FET reaches earlier than SWCNT-FET but at lower degree of ballisticity.

Keywords: CNT-FET, Ballistic, Optimization, DWCNT, SWCNT

**1. Introduction:** Since late 20s, one of the most important aspects of modern electronic research is to scaling down devices to nano-scale. These nano-scaled devices lead to the better performance along with the ability to store large volume of information in very small equipment. Carbon Nanotube (CNT) has the potential to improve the performance of these nanotransistors for which they are good choices for fabrication and optimization of these nanodevices. Single-walled carbon nanotubes are one-dimensional solids [1]. The structure of a Single Walled CNT can be conceptualized by wrapping graphene into a seamless cylinder. On the other hand, Double-walled carbon nanotubes are coaxial nanostructures composed of two single-walled carbon nanotubes, one nested in another [2]. Due to its excellent electrical properties, carbon nanotube is highly attractive for nanoelectronic applications. The band structure of CNT can be tuned to be direct which enables optical emission [3]. Moreover, CNTs are highly resistant to electromigration. Most of the up-to-date CNTFETs operate like non-conventional Schottky barrier transistors [4] which results in quite different device and scaling behaviors from the MOSFET like transistors [5]. Nanotube samples consist of both metallic and semiconducting tubes which is an obstacle to the development for nanoelectronic devices [6]. If the compatibility for silicon device industry is considered, Silicon Nanowire (SiNW) can be a potential candidate [7]. The fact that SiNW blocks can always be semiconducting plays the key role here. On the contrary, CNT-FET has better $I_{ON}/I_{OFF}$ ratio, higher conductance and lower Drain Induced Barrier Lowering (DIBL) than SiNW-FET. Numerous works have been done on comparing SiNW-FET with CNT-FET [8] – [9]

The resistance of short 1-D conductors doesn't depend on the conductor length and composition. It depends on the number of available conduction channels [10]. For a single conduction channel with 100% transparent contacts, the quantum resistance $R_Q$ is given by $R_Q = h/e^2 \sim 26K\Omega$

This is a universal constant. This activity is known as the ballistic conduction. During ballistic transport, electrons travel with no scattering effect between terminals. A model for ballistic nanotransistor had been developed previously [11]. This model was mainly developed for conventional MOSFETs which has been further extended for CNT-FETs and SiNW-FETs. In this paper, this model has been upgraded to observe the non-ballistic effects on the performance of CNT-FET. Our work is based on the FETToy model which is an open source simulation tool provided online [12]. The role of optical phonon scattering on the *I-V* characteristics has been thoroughly investigated previously. The effects of strain on the *I-V* curves of these FETs have also been discussed [13] - [14].

It is a common assumption that scattering increases the resistance against current flow and thus decreases the drain (output) current. From the physical structure of a simple CNT, it is clear that increase in diameter would certainly decrease the effect of scattering. Thus if the term degree of ballisticity is introduced, which is the ratio of current of a CNT-FET with phonon scattering to that without considering phonon scattering, it is expected that this degree of ballisticity will decrease with the increase in diameter. But further investigation which has been carried out in this work shows that this improvement in degree of ballisticity is possible only upto a certain diameter. Decrease in effective drain voltage due the higher diameter plays an important role here. Drain voltage is inversely proportional to the CNT diameter, is reduced which in turn decreases the output current. So there is a trade-off between the degree of ballisticity and CNT diameter. In this work, an empirical equation for this optimized diameter has been developed.

Overall contribution of this work can be summarized as below:

1. An optimization scheme for CNT-FET with optical phonon scattering has been proposed.
2. An optimized diameter for highest Degree of Ballisticity (DoB) can be found with this technique. Its dependence on Gate and Drain Voltage along with channel length has been observed.
3. A Gaussian curved equation for this optimization technique has been developed. Comparison of the optimized diameter for SWCNT-FET and DWCNT-FET is also done.

**2. Methods:** Schottky barrier FETs are considered in our work having metal-semiconductor Schottky barrier at both the source-channel and drain-channel contacts instead of typical MOS-like structures. Numerical simulation using Non-Equilibrium Green's Function Formalism and Boltzmann Transport Equation has been used in this work. The structure is Gate-All-Around (GAA) i.e. the entire circumference is gated for improved performance.

The theory of ballistic nanotransistors [8] implies that the application of electric field between the drain and the source of a nanotransistor would induce a non-equilibrium mobile charge in the channel according to [15]

$$\Delta Q = (N_S + N_D - N_0). \tag{1}$$

Here, $N_S$, $N_D$ and $N_0$ are density of the positive charges filled by the source, density of the negative charges filled by the drain and equilibrium charge densities respectively. These densities are determined by the Fermi-Dirac probability distribution as

$$N_S = \frac{1}{2}\int_{-\infty}^{+\infty} D(E)f(E - U_{SF})dE. \tag{2}$$

$$N_D = \frac{1}{2}\int_{-\infty}^{+\infty} D(E)f(E - U_{DF})dE. \tag{3}$$

$$N_0 = \int_{-\infty}^{+\infty} D(E)f(E - E_F)dE. \tag{4}$$

Here, $D(E)$ is the density of states contributed by the lower sub-band, $E_F$ is the Fermi level, $f$ is the Fermi probability function and $E$ represents the energy levels per nanotube unit length whereas $U_{SF}$ and $U_{DF}$ are the potentials induced by terminal voltages at source and drain respectively as defined by

$$U_{SF} = E_F - qV_{SC}. \tag{5}$$

$$U_{DF} = E_F - qV_{SC} - qV_{DS}. \tag{6}$$

Where $V_{SC}$ is the self-consistent potential and $V_{DS}$ is the induced voltage between drain and source. The concept of self-consistent potential illustrates that the mobile charge potential of nanotransistors can be tuned by terminal voltages. The drain current induced by the transport of non-equilibrium charges across the channel can be calculated by using the Fermi-Dirac statistics represented as [15]

$$I_{DS} = \frac{2qkT}{\pi\hbar}\left[F_0\left(\frac{U_{SF}}{kT}\right) - F_0\left(\frac{U_{DF}}{kT}\right)\right]. \tag{7}$$

Here, $F_0$ is the Fermi integral of order 0, $k$ is the Boltzmann constant, $\hbar$ is the reduced Planck constant ($\hbar = \frac{h}{2\pi}$) and T is the temperature.

Optical phonon scattering is one of the key non-ballistic effects that hamper the FET performance. If the mean free path of optical phonon scattering is smaller than the channel length, which is usually the case for traditional devices, the output performance gets degraded severely due to this scattering effect. To improve this, nano-scale devices can be a great choice where the channel length happens to be smaller. Smaller channel length means that the mean free path of optical phonon scattering is larger than the channel length leading to the lower probability of getting scattering before passing through the channel region. As a result, there is a significant improvement in *I-V* characteristics. Previous works [15] have emphasized on the development of equations for the transmission coefficient due to phonon scattering. The effective phonon scattering mean free path in semiconducting nanotubes can be computed by equations [8-11].

$$\frac{1}{l_{SC}(V_x)} = \frac{1}{l_{ap}}\left[1 - \frac{1}{1+e^{\frac{(E_F - V_{SC} + qV_x)}{KT}}}\right]$$

$$+ \frac{1}{l_{op}}\left[1 - \frac{1}{1+e^{\frac{(E_F - qV_{SC} - \hbar\omega_{op} + qV_x)}{KT}}}\right]. \tag{8}$$

$$T_S = \frac{l_{SC}(0)}{l_{SC}(0)+L}. \tag{9}$$

$$T_D = \frac{l_{SC}(V_{DSeff})}{l_{SC}(V_{DSeff})+L}. \tag{10}$$

$$I_{DSP} = \frac{2qkT}{\pi\hbar}\left\{T_S \ln\left(1 + e^{\frac{E_F - qV_{SC}}{KT}}\right)\right.$$

$$\left. - T_D \ln\left(1 + e^{\frac{E_F - q(V_{SC} + V_{DSeff})}{KT}}\right)\right\}. \tag{11}$$

Where $l_{ap}$ = 500 nm is a typical acoustic phonon scattering mean free path value while $l_{op}$ =15 nm is a typical optical phonon scattering mean free path and $\hbar\omega_{op}$ is a typical OP energy. It can be noticed that at low carrier energy (e.g. < 0.15 eV), the acoustic scattering dominates, while the optical scattering is more important at high kinetic energy. Here, $T_S$ and $T_D$ are scattering coefficients at source and drain respectively whereas $I_{DSP}$ corresponds to the current due to phonon scattering.

The equation for effective channel potential drop due to optical phonon scattering can be written as [15]

$$V_{DS_{eff}} = \frac{L}{L + \frac{d}{d_0}\lambda_{op}} V_{DS}. \tag{12}$$

Where $\lambda_{op}$ is the optical mean free path, $d$ is the diameter of the tube, $d_0$ is the reference diameter and $L$ is the length of the channel. If we consider the applied drain bias as $V_{DS_{eff}}$ instead of $V_{DS}$, the *I-V* characteristics with optical phonon scattering is found which has been illustrated in this work.

In our work, we have introduced a new term, the 'Degree of Ballisticity' (DoB). It is simply the ratio of drain current with optical phonon scattering to drain current without optical phonon scattering. It is expected that this value will be less than 1. The closer this value to the 1, the better the device is. According to the theory, higher diameter would mean that optical phonon will pass through the channel without getting scattered as much as it does for same channel length with smaller diameter. It is a common assumption that the DoB would increase for higher diameter.

In general, this phenomenon is true for smaller diameters. For higher diameters, there is another effect that dominates. This is the reduction in effective drain voltage due to higher diameter. Lower effective drain voltage would decrease the output current, that is it would decrease the DoB. As a result, there are two opposite effects.

For smaller diameter, the increment due to diameter increase dominates over drain voltage reduction. But for higher diameter, drain voltage reduction dominates. So there must be a trade-off between these two effects. The primary aim of this work is to optimize the diameter dimension as well as biasing characteristics so that the best possible output (degree of ballisticity) can be obtained for a specific dimension and bias condition.

$$Degree\ of\ Ballisticity, R = \frac{I_D\ (non-ballstic)}{I_D\ (ballstic)}. \tag{13}$$

Here, $I_D$ (non-ballistic) corresponds to $I_{DSP}$ which is given by (11). The basic concept of the proposed optimization technique is depicted in Fig. 1. From the figure, it can be assumed that the optimized diameter to give the best performance should be the halfway through the diameter range.

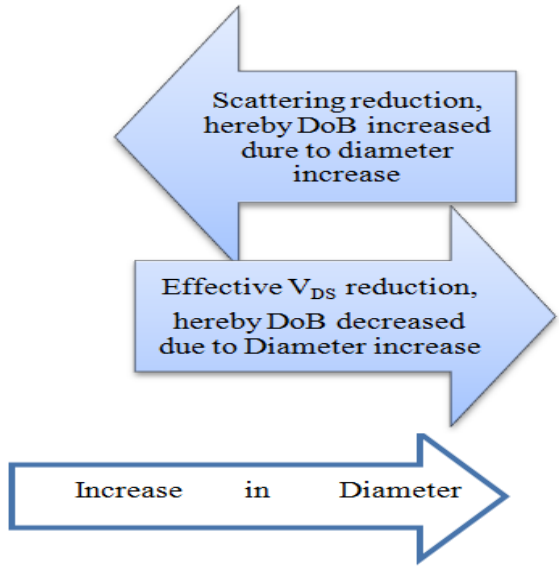

Figure 1 Proposed diameter optimization technique for optical phonon scattering effect

**3. Simulation Results:** As mentioned earlier, a trade-off between output drain voltage and increase in diameter effect on degree of ballisticity has been worked out here. These two opposite effects will make the degree of ballisticity vs diameter graph a dome shaped curve where there is a peak value corresponding to the highest degree of ballisticity for a certain diameter. This diameter can be considered as the optimum diameter.

From the simulation, the dome shaped curve has been obtained (Fig. 2). At first, the curve has been simulated for 1V drain bias and 0.6V gate bias and it is observed that for diameter < 1.1 nm, increment in DoB due to diameter increase dominates; while for > 1.3 nm, the drain voltage reduction effect dominates. There happens to be an optimum diameter at 1.25 nm.

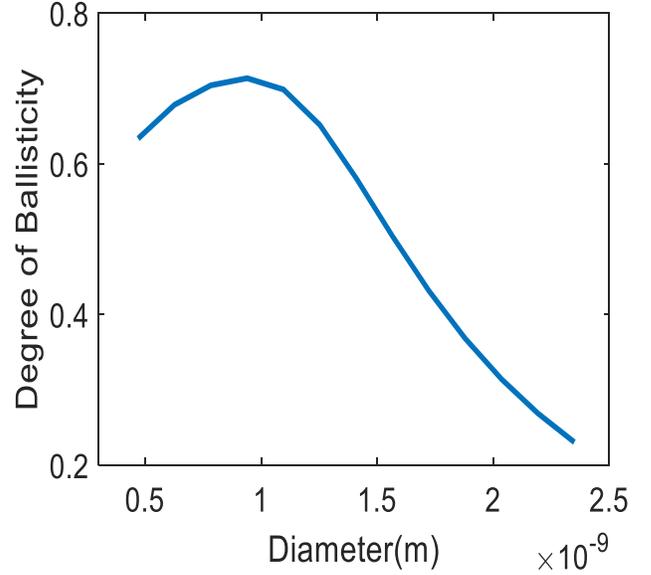

Figure 2. Dependence of degree of ballisticity with tube diameters.

Effective drain voltage is different for different applied bias. As the effective $V_{DS}$ is one of the two key factors effecting the dome-shaped curve it is expected that optimum diameter will depend on this applied $V_{DS}$. Moreover, applied gate bias $V_{GS}$ should also affect this optimum diameter ($D_{opt}$).

For the typical MOSFET structure, it is known that the output current depends on the difference between the $V_{DS}$ and $V_{GS}$. With a view to finding a relation between the optimum diameter and ($V_{DS} - V_{GS}$), the latter was varied to observe their effects on optimum diameter and degree of ballisticity, keeping the channel length (L) same (Fig. 3). For the variation of ($V_{DS} - V_{GS}$), $V_{GS}$ was varied first keeping $V_{DS}$ fixed (Fig. 3a) and vice-versa (Fig. 3b). Thus, for different values of ($V_{DS} - V_{GS}$), their effects on optimum diameter and degree of ballisticity have been observed in table 1.

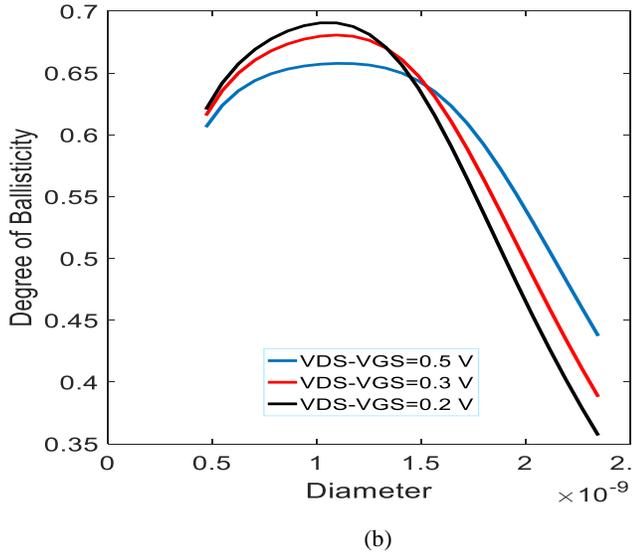
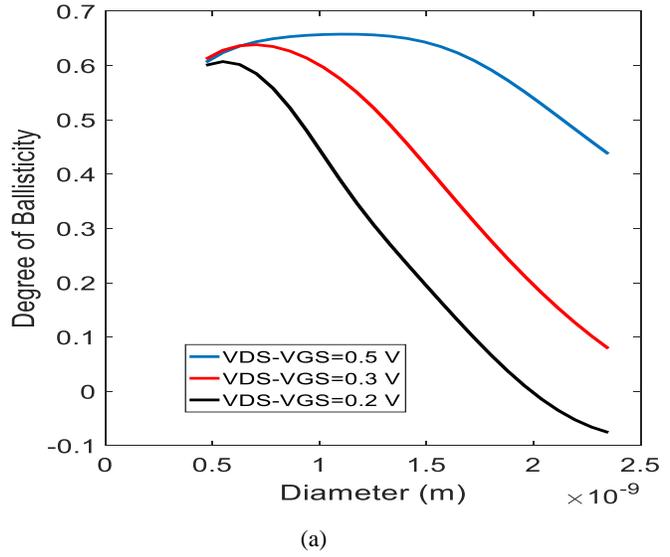

(b)     (a)

Figure 3. Effect of diameter on degree of ballisticity for different ($V_{DS} - V_{GS}$): (a) Gate voltage variation (b) Drain voltage variation.

TABLE I. EFFECT OF ($V_{DS}$ -$V_{GS}$) ON OPTIMUM DIAMETER AND DEGREE OF BALLISTICITY

| ($V_{DS}-V_{GS}$) (V) | Variable Voltage | Optimum Diameter (nm) | Degree of Ballisticity at Optimum Diameter |
|---|---|---|---|
| 0.5 | Drain | 1.175 | 0.6578 |
| 0.3 | Drain | 1.096 | 0.6807 |
| 0.2 | Drain | 1.080 | 0.6905 |
| 0.3 | Gate | 0.624 | 0.6384 |
| 0.2 | Gate | 0.548 | 0.6073 |

Table I shows the changes in the optimized diameter as well as the DoB. Here, the optimum diameter increases with ($V_{DS} - V_{GS}$). The fact is that the higher the difference between drain voltage and gate voltage, the more current will pass through drain to source. As a result, number of scattering will be increased for higher $V_{DS}$-$V_{GS}$. Consequently, phonon scattering due to diameter increase will predominate over the drain voltage reduction up to a larger limit of diameter. Moreover, there is another fact that can be observed from Fig. 3. Higher $V_{DS}$-$V_{GS}$ creates higher DoB for higher diameter (more than optimized diameter. However, higher $V_{DS}$-$V_{GS}$ reduces DoB for smaller (less than optimized diameter) diameter. It can be explained by the phenomenon that for higher diameter reduces the effective $V_{DS}$ from (12) to a comparatively lower value for smaller $V_{DS}$-$V_{GS}$, thereby DoB is decreased.

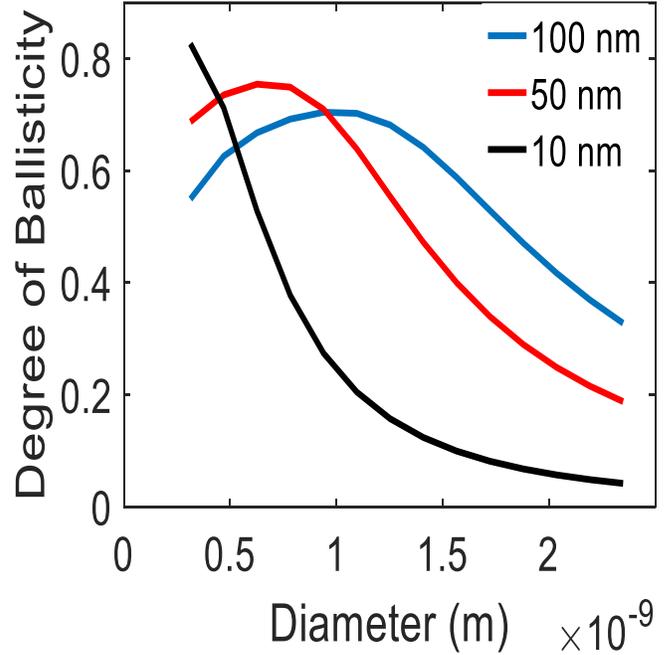

Figure 4. Effect of diameter on degree of ballisticity for different channel length.

Channel Length variation is another important factor that affects the performance of the device. Decreasing length improves the ballistic performance, as the theory suggests that the phonon scattering mean free path gets larger than the nanotube length in that case. As mentioned earlier, increase in length causes the degree of ballisticity to be reduced. So, higher diameter is needed to reach the optimum point for longer channel (Fig. 4).

For finding optimum diameter for highest degree of ballisticity, the degree of ballisticity has to be maximized with respect to diameter i.e.

$$\frac{d(R)}{d(d)} = 0 \tag{14}$$

From the above discussion, it can be assumed that thedegree of ballisticity can be written as,

$$R = aL\,[(V_{GS} - V_{Th})V_{DS} - V_{DS}^2/2] + b \tag{15}$$

Where a and b are parameters on which the DoB depends. It has been discussed previously that the optimum diameter increases as $(V_{DS} - V_{GS})$ increases and it also increases with channel length L. As a result, our proposed equation for optimum diameter can be formulated as

$$D_{optimum} = cL(V_{DS} - V_{GS}) + d \tag{16}$$

Curve fitting tool of MATLAB has been used to develop an empirical equation for determining the relationship betwwen diameter and degree of ballisticity. The characteristics being a dome shaped curve, it can be considered as a gaussian curve. Hence, a gaussian equation can be proposed as in (17).

$$R = pe^{-(\frac{x-q}{c})^2} \tag{17}$$

Where the fitted parameters are:

p = 0.6834,
q = 0.0274 and
c = 2.411.

Here, q is the optimum point i.e. the desired optimum diameter $D_{optimum}$.

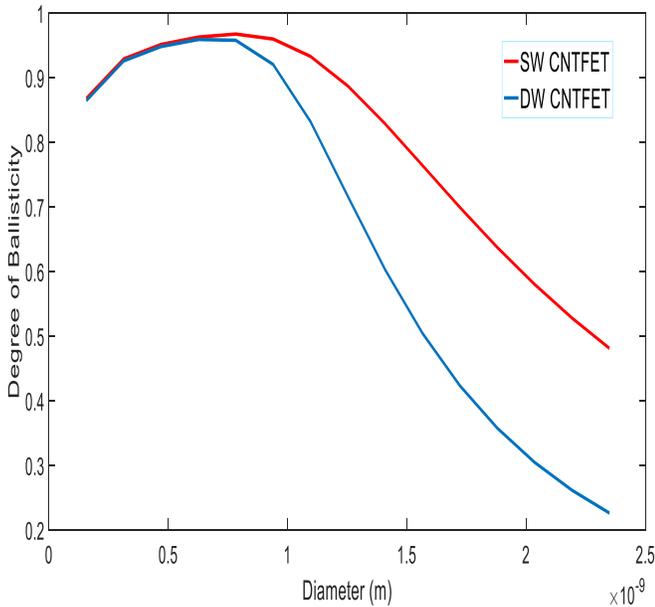

Figure 5. Optimum diameter comparison between SW CNTFET and DW CNTFET.

This empirical equations can be extended to DW CNT-FET. Double walled CNT has both inner wall and outer wall where inter wall coupling between both walls lead to the higher optical phonon scattering. As a result, degree of ballisticity for DW CNTFET will be less than that of SW CNT-FET. However, Optimum diameter for DW CNT-FET reaches earlier (0.4697 nm) than SW CNT-FET (0.6264 nm) but at lower degree of ballisticity (Fig. 5).

In the case of DWCNT-FET, this optimum diameter refers to the optimum diameter of the outer wall. In this work, inter wall distance is considered same, i.e. the diameters of both walls have been increased by same amount to find the optimum diameter.

Our proposed technique is developed theoretically. Experimental validity can be tested in future. Moreover, dependence of highest Degree of Ballisticity on other parameters such as the types of CNT (Metallic or semiconducting), chirality, gate-oxide thickness can be analyzed. Fitting parameters a, b, c, d, p, and q can be derived more accurately and can be matched with experimental result.

**4. Conclusion:** To sum up, an optimization of diameter for achieving the maximum ballisticity has been proposed for CNT-FET corresponding to the relationship between diameter and degree of ballisticity. It has also been found that degree of ballisticity along with the optimum diameter increases with the increase of $(V_{DS} - V_{GS})$. The increase of channel length, however, degrades the ballistic performance demanding a higher diameter to reach the optimum point. An equation using the Gaussian distribution has been proposed to achieve the optimization. Moreover, the optimum diameter for highest degree of ballisticity occurs at a lower value for DWCNT than its SW counterpart.

**5. Acknowledgments:** The basic FETToy 2.0 model which is used in this work is available online [12].